\documentclass[a4paper,fleqn,usenatbib]{mnras}
\usepackage{newtxtext,newtxmath}

\usepackage[T1]{fontenc}
\usepackage{ae,aecompl}

\usepackage{graphicx}	
\usepackage{amsmath}	
\usepackage{amssymb}	
\usepackage{multirow}
\usepackage[version=4]{mhchem}
\usepackage{pdflscape}
\usepackage{gensymb}
\usepackage{hyperref}

\title[The changing secondary eclipse depth of WASP-12b]{Storms or Systematics? The changing secondary eclipse depth of WASP-12b}

\author[M. J. Hooton et al.]{
Matthew J. Hooton,$^{1}$\thanks{E-mail: mhooton01@qub.ac.uk}
Ernst J. W. de Mooij,$^{2}$
Christopher A. Watson,$^{1}$ \newauthor
Neale P. Gibson,$^{1}$
Francisco J. Galindo-Guil,$^{3,4,5}$
Rosa Clavero$^{6,4}$
and \newauthor
Stephanie R. Merritt$^{1}$
\\
$^{1}$Astrophysics Research Centre, 
School of Mathematics and Physics,
Queen's University Belfast,
Belfast BT7 1NN, UK\\
$^{2}$School of Physical Sciences, 
Dublin City University,
Glasnevin,
Dublin 9, Ireland\\
$^{3}$Nordic Optical Telescope,
Nordic Optical Telescope Apartado 474E-38700
Santa Cruz de La Palma, Santa Cruz de Tenerife, Spain\\
$^{4}$Isaac Newton Group of Telescopes, 
Apartado de Correos 321, 38700 Santa
Cruz de La Palma, Spain\\
$^{5}$Depto. de  Astrof\'isica,
Centro  de  Astrobiolog\'ia  (INTA-CSIC), ESAC
campus,Camino  Bajo  del  Castillo  s/n.  28692  Villanueva de la\\
Ca\~nada, Madrid, Spain\\
$^{6}$Instituto de Astrof\'isica de Canarias, 
C/ Via L\'actea, s/n E38205, Spain
}

\date{Accepted 2019 April 2. Received 2019 March 29; in original form 2019 February 13}

\pubyear{2018}

\begin{document}
\label{firstpage}
\pagerange{\pageref{firstpage}--\pageref{lastpage}}
\maketitle

\begin{abstract}

WASP-12b is one of the most well-studied transiting exoplanets, as its highly-inflated radius and its 1.1 day orbit around a G0-type star make it an excellent target for atmospheric categorisation through observation during its secondary eclipse. We present two new secondary eclipse observations of WASP-12b, acquired a year apart with the Wide Field Camera on the Isaac Newton Telescope (INT) and the IO:O instrument on the Liverpool Telescope (LT). These observations were conducted in the $i^\prime$-band, a window expected to be dominated by TiO features if present in appreciable quantities in the upper atmosphere. We measured eclipse depths that disagree with each other by $\sim$3$\sigma$ (0.97 $\pm$ 0.14 mmag on the INT and 0.44 $\pm$ 0.21 mmag on the LT), a result that is mirrored in previous $z^\prime$-band secondary eclipse measurements for WASP-12b. We explore explanations for these disagreements, including systematic errors and variable thermal emission in the dayside atmosphere of WASP-12b caused by temperature changes of a few hundred Kelvin: a possibility we cannot rule out from our analysis. Full-phase curves observed with \textit{TESS} and \textit{CHEOPS} have the potential to detect similar atmospheric variability for WASP-12b and other optimal targets, and a strategic, multi-telescope approach to future ground-based secondary eclipse observations is required to discriminate between explanations involving storms and systematics.

\end{abstract}




\section{Introduction}

Secondary eclipse observations are an important tool to characterise the atmospheres of transiting hot Jupiters. The vast majority of secondary eclipse studies have been conducted in the near-infrared (NIR), at wavelengths of 1.1 $\mathrm{\mu}$m and longer \citep[e.g.][]{2005ApJ...626..523C,2009A&A...493L..35D,2013Icar..225..432S}. At optical wavelengths and shorter, the flux from hot Jupiters due to thermal emission drops off sharply in contrast to their host stars. With current instrumentation, individual secondary eclipse signals are not detectable in the optical for all but the most heavily irradiated exoplanets, though the precision afforded by the NIR instruments mounted on the \textit{James Webb Space Telescope} is expected to be able to detect considerably smaller signals at these wavelengths than is currently possible \citep[e.g.][]{2018PASP..130k4402B}. 

For the handful of exoplanets with sufficiently large signals, secondary eclipse observations at red-optical wavelengths are a rich source of information about their compositions and temperature structures. The $i^\prime$- and $z^\prime$-bands (with coverage of $\sim$0.7-1.1 $\mathrm{\mu}$m) contain prominent titanium oxide (TiO) and vanadium oxide (VO) features, which should dominate the spectra of hot Jupiters if present in appreciable quantities in their upper atmospheres. TiO and VO have long been predicted to be significant opacity sources that could give rise to temperature inversions in the most highly-irradiated exoplanets \citep[e.g.][]{2003ApJ...594.1011H,2008ApJ...678.1419F}. Despite numerous searches \citep[e.g.][]{2013MNRAS.436.2956S,2015ApJ...806..146H}, high significance TiO detections have only been reported more recently. \citet{2017AJ....154..221N} and \citep{2017Natur.549..238S} reported detections of TiO at high resolution in the WASP-33b dayside and at low resolution at the WASP-19b terminator respectively. In the case of WASP-19b, this contradicts non-detections by \citet{2013MNRAS.434.3252H} and \citet{2019MNRAS.482.2065E} in the same wavelength range. Tentative reports of TiO and VO in the transmission spectrum of WASP-121b at NIR wavelengths \citep{2016ApJ...822L...4E} were complicated by a recent study that found evidence for VO, but not TiO, in its optical transmission spectrum \citep{2018AJ....156..283E}.

As observations of secondary eclipses in the $i^\prime$- and $z^\prime$-bands probe the spectral energy distributions of exoplanets at wavelengths that are shortward of their peaks, the thermal emission is very sensitive to the precise temperature of the emitting layer. \citet{2012ApJ...758...36M} predicted that for many hot Jupiters, thermal emission measurements at these wavelengths are good discriminators between carbon-rich and carbon-poor exoplanets. Whilst secondary eclipse observations conducted at $z^\prime$-band wavelengths have been published for 9 exoplanets\footnote{We include $i^\prime$- and $z^\prime$-band observations with narrow-band filters that fall within their respective wavelength coverages, but not observations with wider bandpasses that overlap with these filters \citep[e.g.][as well as various detections using \textit{Kepler} and \textit{TESS} data.]{2013A&A...553A..49A}} \citep{2009A&A...493L..31S,2010ApJ...716L..36L,2011MNRAS.416.2096S,2012ApJS..201...36B,2012A&A...542A...4G,2013A&A...552A...2L,2013ApJ...774..118Z,2013MNRAS.435.2268F,2014A&A...567A...8C,2016MNRAS.458.4025D,2017Natur.546..514G,2018MNRAS.474.2334D}, only two studies have reported secondary eclipse observations at $i^\prime$-band wavelengths \citep{2013MNRAS.436....2M,2014A&A...567A...8C}, with the most confident detection reported at 3.7$\sigma$.

WASP-12b \citep{2009ApJ...693.1920H} is the subject of many disagreements regarding the classification of its atmosphere. Its 1.09 day orbit around a G0-type star makes WASP-12b one of the most heavily irradiated known exoplanets, resulting in a dayside temperature of $\sim$3000 K. This, as well as its highly-inflated size, results in some of the largest eclipse depths for any known exoplanet, with secondary eclipse observations of WASP-12b having been carried out across the NIR. Its large scale height also makes it an excellent target to study its terminator using transmission spectroscopy. \citet{2012ApJ...758...36M} used CFHT and Spitzer secondary eclipse observations \citep{2011AJ....141...30C,2011ApJ...727..125C} to infer that WASP-12b has a high C/O ratio. Later observations cast doubt on this assessment, as \citet{2014ApJ...791...36S} and \citet{2015ApJ...814...66K} found that the emission and transmission spectra of WASP-12b could be well fit by models without a high C/O ratio. \citet{2017ApJ...847L...2B} measured a geometric albedo ($A_\mathrm{g}$) < 0.064 for WASP-12b at optical and near-ultraviolet (NUV) wavelengths. This suggests that WASP-12b is amongst the darkest-known exoplanets, which is contrary to transmission spectroscopy studies that predict significant scattering from high-altitude aerosols \citep[e.g.][]{2013MNRAS.436.2956S,2017ApJ...834...50B}.

In this paper, we present two $i^\prime$-band secondary eclipse observations of WASP-12b that were conducted a year apart. The measured depths significantly disagree, mirroring the results of similar observations conducted at $z^\prime$-band wavelengths \citep{2010ApJ...716L..36L,2013MNRAS.435.2268F}. We also present an analysis of all reported WASP-12b secondary eclipse depths, in addition to a discussion of whether this could arise due to systematic errors or detectable variability in the WASP-12b dayside. In section \ref{sec:observations}, we describe our observations and our data reduction; in section \ref{sec:analysis}, we describe how we decorrelated the light curves before presenting our results, which show a significant disagreement in the eclipse depths. In section \ref{sec:discussion}, we discuss the possible reasons for the disagreement and in section \ref{sec:conclusions}, we present our conclusions and outline possible future observational strategies to discriminate between these scenarios.

\section{Observations \& Data Reduction}
\label{sec:observations}

\subsection{Isaac Newton Telescope (INT) Observations}
\label{sec:int}

We observed one secondary eclipse of WASP-12b on 2017 January 15 using the Wide Field Camera \citep[WFC;][]{1998IEEES..16...20I} with the Sloan Gunn $i^\prime$ filter\footnote{Whilst we adopt a convention of referring to both filters used in our observations as $i^\prime$-band filters, the WFC filter has a slightly redder response function. This is indicated in Figure \ref{models}, and taken account of in the contamination corrections described in Appendix \ref{appendix}.} on the Isaac Newton Telescope (INT) on La Palma, using a strategy broadly similar to the one adopted in \citet{2018ApJ...869L..25H}. The WFC consists of a mosaic of 4 CCDs, each with a plate scale of 0$.\!\!^{\prime\prime}$33 per pixel. We only used the central CCD (CCD4), resulting in a $\sim$22$.\!\!^\prime$7 by 11$.\!\!^\prime$4 field-of-view. During the observation, the Moon had an average angular separation of 64$\degree$ from WASP-12 and an average illumination of 85\%. The seeing was poor and varied between 1-3$^{\prime\prime}$. The observation began at 21:07 UT and lasted $\sim$ 7.9 hours. During this time, we obtained 264 frames with an exposure time of 60 seconds and readout time of $\sim$29 seconds. WASP-12b was fully or partially occulted by its host star for 103 of these frames. At the very end of the observation, 22 frames were discarded as the low altitude of WASP-12 resulted in  vignetting from the lower shutter of the INT.

The INT was defocused during our observation. This strategy minimises flat-fielding errors, reduces overheads, and makes the point spread functions (PSFs) more robust against the variable seeing that was observed. The resulting donut-shaped PSFs had diameters of 27 pixels (9$^{\prime\prime}$). As the auto-guider for the INT requires the telescope to be in focus, we used a custom code to guide the telescope.  This accounts for any drift in the telescope by using the science frames. The total drift was less than 9 pixels (3$^{\prime\prime}$) throughout the night, with a root mean square (RMS) of 1.9 pixels. We ensured that WASP-12 and the other comparison stars were positioned on well-behaved parts of the detector. 

The frames were reduced using a custom IDL pipeline written by the lead author. Each of the images was bias-subtracted row-by-row using the mean of the overscan regions at each edge of the CCD, followed by a full-frame subtraction to account for further artefacts present in the bias frames. The row-by-row subtraction addresses an issue where the bias level in the WFC CCDs periodically jumps to different values--a phenomenon that was observed in 43 of the frames. Each of the frames was flat-fielded using twilight flats. 

A fringing pattern with an amplitude of $\sim$5\% of the background flux was present in the science frames throughout the observation. Although the position of the fringes on the WFC detector is stable with time, small movements in the target and comparison stars across the detector can introduce systematic errors into the light curves if they cross any fringes. We created a map of the fringing pattern by acquiring dithered frames on a blank field with an exposure time of 60 seconds, and combined the frames using a clipped median of each pixel.
    
As the intensity of the fringing pattern in the data varied with time, we needed to scale the fringing map accordingly before subtracting it from each frame. We explored a couple of methods to do this in a robust and consistent way. Firstly, we linearly fitted the whole fringing map to the whole of each science frame with the stars masked. This produced well-corrected science frames for the majority of the time-series, but failed to robustly remove the fringes for frames acquired between phases 0.37-0.42, where contamination from the Moon caused a gradient across the frames. Instead, we selected a small region of the science frames with a good range in fringe amplitudes, and used this region to fit for the proper offset and scaling for the fringe frame using a linear fit between the fringe map and the science frame, taking care to mask any stars in the region. This approach was significantly less sensitive to the gradient across the Moon-contaminated science frames and robustly removed the fringes for the entire time-series.

Finally, we fit and subtracted a second-order 2-dimensional polynomial from each of the science frames with the stars masked. This step was performed to correct the small trends in background flux that were a function of their position on the CCD \citep[e.g.][]{2011A&A...528A..49D}, including contamination from the Moon that affected $\sim$40 out-of-eclipse frames. A small gradient was still present in the Moon-contaminated frames that was not removed by higher-order polynomial fits. For six frames that centred on WASP-12 crossing meridian, a more complex trend was visible across the CCD, which our polynomial fit was unable to successfully model. For this reason, we excluded these frames from our analysis.

Five bright comparison stars within 5$^\prime$ of WASP-12b (with $V_\mathrm{mag}$ = 11.52, 11.62, 11.74, 12.22 and 13.08) were selected to normalise the target light curve, as stars further away typically introduced large systematic errors into the time-series due to the position-dependent effects of the Moon-contamination on the detector.

Our method of defocusing the telescope resulted in two nearby background stars (2MASS06303343+2940255 and 2MASS06303255+2940301; referred to hereafter as 2M063033 and 2M063032, respectively) blending with the WASP-12 PSF. The target and comparison stars were centred by picking a bright star in a clear area of the detector, computing the mean $x$ and $y$ positions of the pixels with flux $5\sigma$ above the background level in every frame and adding offsets to calculate the positions of the other stars.

We performed aperture photometry using the APER procedure from the IDL Astronomy User's Library\footnote{idlastro.gsfc.nasa.gov} on each of the five comparison stars with a radius of 44 pixels. This was sufficiently large to capture $\gg$ 90\% of the flux from both background stars so that their contributions to the flux could be quantified and accounted for (see Table \ref{tab:contaminants}), and larger aperture sizes resulted in an increased RMS due to the addition of more background flux. The residual sky background for each star was computed in a sky annulus outside the target aperture, which had inner radius of 50 pixels and an outer radius 69 pixels. A range of inner and outer radii were tested, but the changes had negligible effects on the measured background flux.


\subsection{Liverpool Telescope (LT) Observations}
\label{sec:lt}

We observed one secondary eclipse of WASP-12b on 2018 January 20 using the IO:O instrument with the SDSS-i filter on the Liverpool Telescope \citep[LT; ][]{2004SPIE.5489..679S} on La Palma. IO:O in the standard 2x2 binning mode is a 2048x4056 pixel e2v detector with a plate scale of 0$.\!\!^{\prime\prime}$30 per pixel, giving us a field-of-view of $\sim$10$.\!\!^\prime$24 by 10$.\!\!^\prime$28. During the observation, the Moon had an average angular separation of 111$\degree$ from WASP-12 and an average illumination of 13\%. The seeing was also poor and varied between 1-3.5$^{\prime\prime}$. The observation began at 23:53 UT and lasted $\sim$ 4.7 hours. During this time, we obtained 393 frames with an exposure time of 40 seconds and readout time of 18.5 seconds. WASP-12b was fully or partially occulted by its host star for 185 of these frames.

The LT was also defocused during the observation. The resulting donut-shaped PSFs had a diameter of 10 pixels (3$^{\prime\prime}$). The auto-guider for the LT maintains its own focus independently of IO:O, and was hence suitable for use for these observations despite our defocused PSFs. A drift of 9 pixels occurred in the 25 frames taken immediately after meridian; the drift throughout the remainder of the observations was $<$ 4 pixels (1.2$^\prime$), with an RMS of 0.8 pixels.

As standard, the science frames are bias-subtracted, overscan-trimmed and flat-fielded using the IO:O pipeline at the observatory. The images were bias-subtracted by subtracting a single value from the entire image based on the overscan regions on either side of the image. These overscan regions were then trimmed, before the frames were each flat-fielded using a master flat constructed from twilight flats. Unlike for the WFC, negligible fringing was observed in the IO:O frames and no correction was performed. When stacking the science frames, a small trend was still visible across the detector after the IO:O reduction, despite the flat-fielding. For this reason, a second-order polynomial was fit to and subtracted from the entirety of each frame individually with the stars masked.

We performed aperture photometry on the same five comparison stars used in the INT reduction using a broadly similar method. As the LT resolves the world coordinate system, this was used to compute the positions of the stars in each frame. We selected target apertures with radii of 54 pixels, and inner and outer radii of 60 and 79 pixels, respectively. 

\section{Analysis}
\label{sec:analysis}

\begin{figure}
\includegraphics{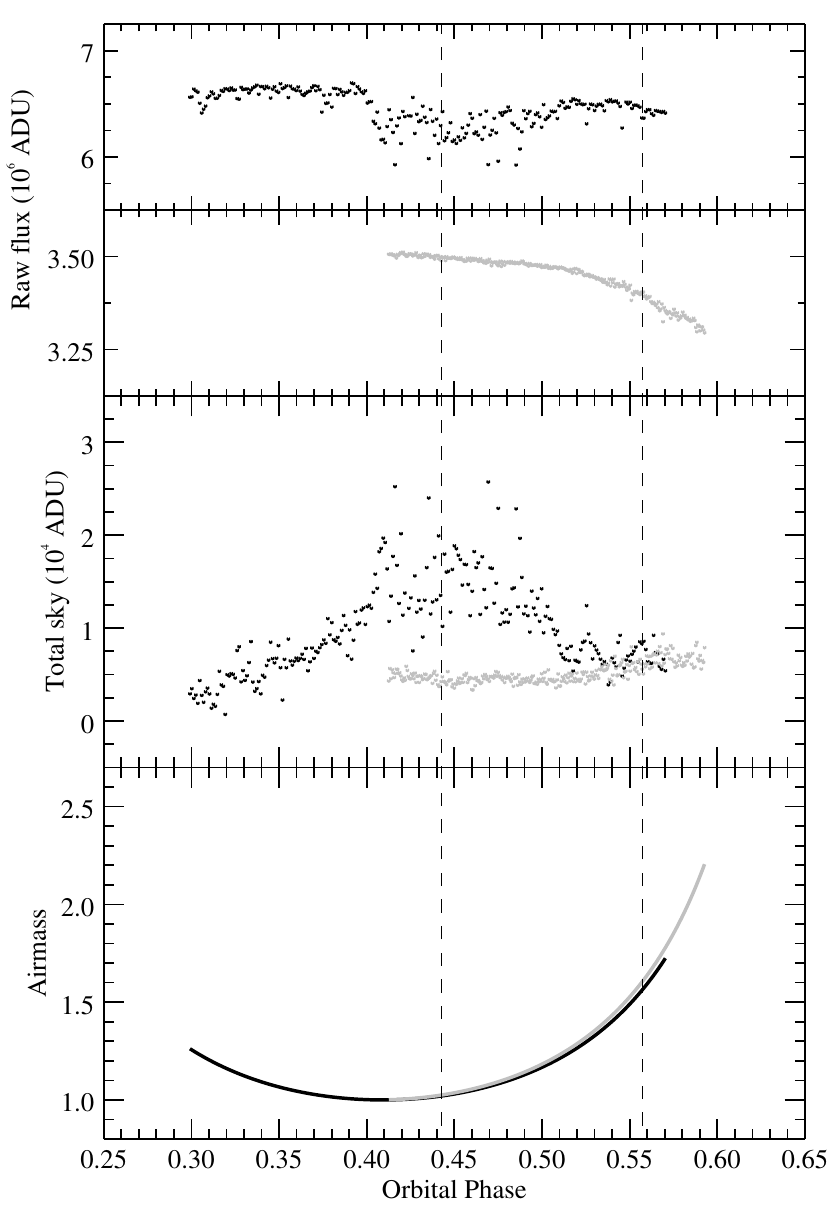}
\caption{Measurements associated with the INT (black points) and LT (grey points) observations. The orbital phases at which ingress begins and egress ends for our best fit MCMC model are marked with dashed lines. Top two panels: raw flux inside the target aperture. Third panel: total flux inside the target sky annulus. Bottom panel: airmass.}
\label{parameters}
\end{figure}

The raw target light curves for WASP-12 are shown in the top two panels of Figure \ref{parameters}. A dip in the raw flux in the INT light curve between phases 0.40 and 0.51, as well as the simultaneous spike in the flux in the sky annulus (see the third panel of \ref{parameters}), corresponds to a time when thin cirrus clouds were likely passing overhead. The flux of a few star-free sections on the detector was measured, and all show same spike in sky brightness that is visible in the sky annuli of the target and reference stars. As we observed correlations between the flux and the sky background for both the INT and LT data, sky background was included as a component in the baseline model for both observations.

\begin{figure}
\includegraphics{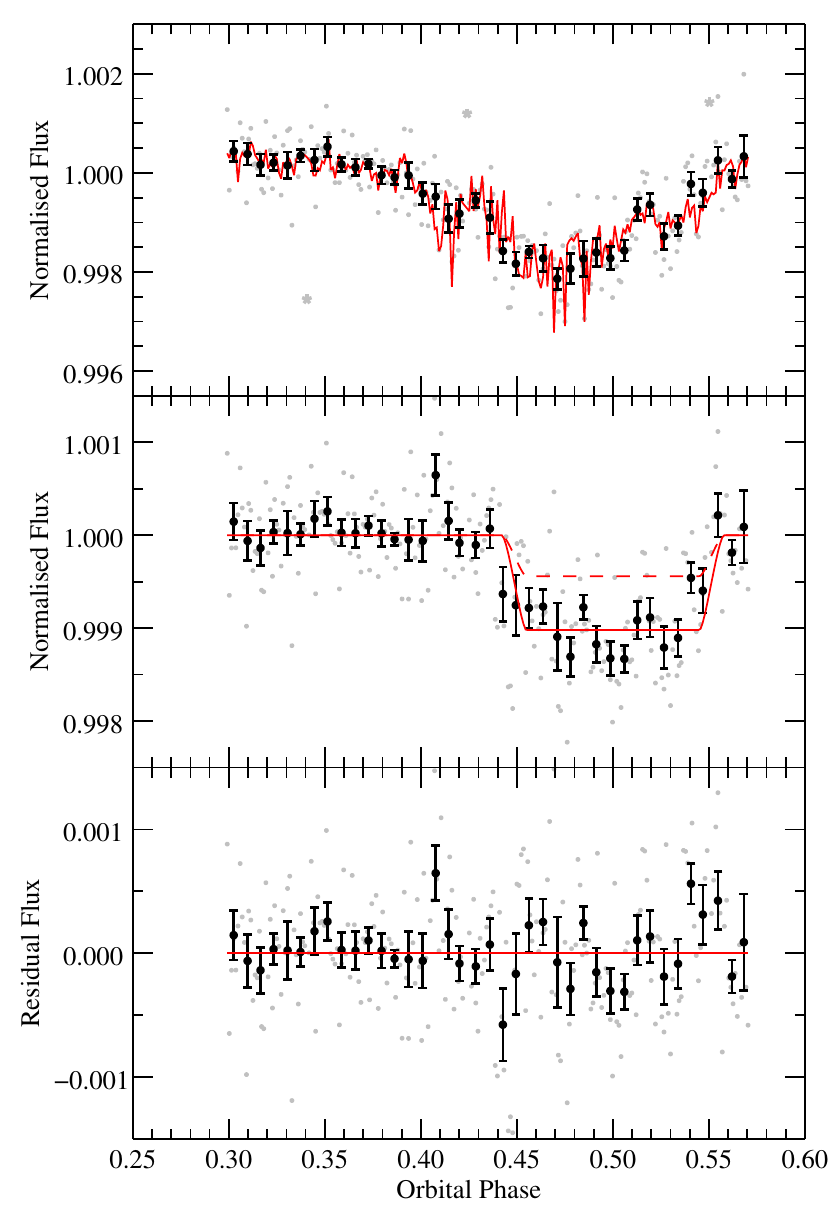}
\caption{INT photometry of WASP-12 acquired on 2017 January 15. Top panel: the light curve unbinned (grey points) and in phase bins of 0.007 (black points) after correcting for flux from the stellar companions and background stars in the target apertureThe best fit MCMC model is shown in red. Points that were removed by clipping at 3$\sigma$ are shown as grey asterisks. Middle panel: the light curve after dividing by the baseline model, along with the best fit eclipse model (solid red line). The best fit eclipse model from the LT observation is indicated with a dashed red line. Bottom panel: the residual flux.
}
\label{INT}
\end{figure}

\begin{figure}
\includegraphics{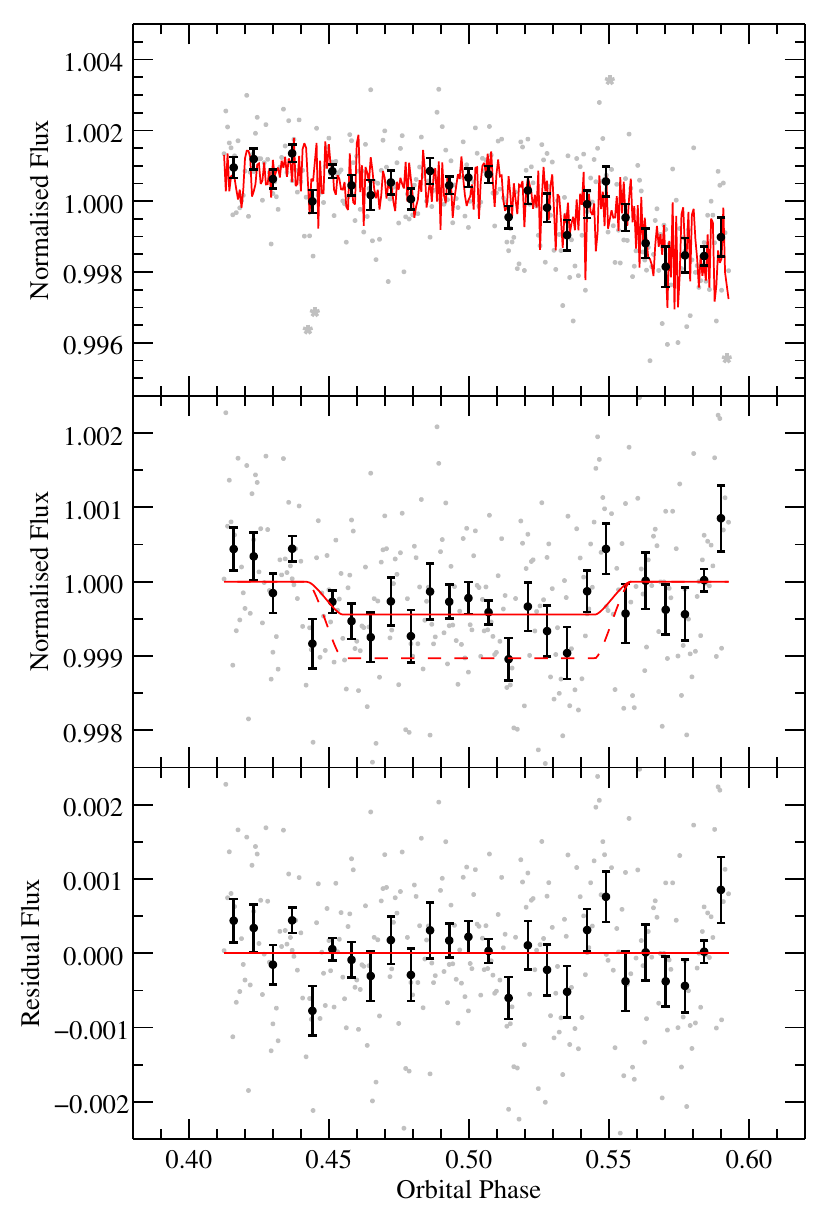}
\caption{The same as Figure \ref{INT} except for LT photometry acquired on 2018 January 20. The dashed red line in the third panel shows the best fit eclipse model for the INT data.}
\label{LT}
\end{figure}

 The normalised light curves for each observation were created by dividing the raw target light curves by the total of the five comparison stars, and dividing through to get a median out-of-eclipse baseline of 1. The comparison stars selected had similar magnitudes and colours to WASP-12, to ensure the SNR was maximised and that no significant colour terms were introduced into the normalised light curve. This step removed the visible correlation with airmass (see the bottom panel of Figure \ref{parameters}) for the INT light curve, but did not fully remove it for the LT light curve. Despite the drift at the start of the LT observation, we did not observe a correlation with the $x$ and $y$ positions of the stars on the detector in either dataset. 

\begin{table}
\centering
 \caption{Parameters of the WASP-12A system. A - \citet{2009ApJ...693.1920H}, B - this work.}
 \label{tab:parameters}
 \begin{tabular}{ccc}
  \hline
  Parameter & Value & Ref.\\
  \hline
  \multicolumn{3}{c}{Stellar Parameters} \\
  $R_*$ ($R_\odot$)  & $1.57\pm 0.07$ & A \\
$T_{*}$ (K) & $6300^{+200}_{-100}$ & A \\
\textbf{log}($g$) & $4.38\pm 0.10$ & A \\
$[$Fe/H$]$ & $0.30^{+0.05}_{-0.15}$ & A \\
\hline
\multicolumn{3}{c}{Planetary Parameters}\\
$R_\mathrm{P}$ ($R_\mathrm{J}$) & $1.79 \pm 0.09$ & A \\
$t_\mathrm{0}$ (MJD) & 54508.4761 $\pm$ 0.0002 & A \\
$P$ (days) & 1.091423 $\pm$ 0.000003 & A \\
$a$ (au) & 0.0229 $\pm$ 0.0008 & A \\
$i$ ($^{\circ}$) & $83.1^{+1.4}_{-1.1}$ & A \\
\hline
\multicolumn{3}{c}{INT Measured Parameters}\\
$F_\mathrm{ecl}$ (mmag) & 0.97 $\pm$ 0.14 & B \\
$\sigma_\mathrm{w}$ & (4.7 $\pm$ 3.4) x 10$^{-4}$ & B \\
$\sigma_\mathrm{r}$ & (1.12 $\pm$ 0.40) x 10$^{-4}$ & B \\
$T_\mathrm{B}$ (K) & $3412^{+85}_{-93}$  & B \\
\hline
\multicolumn{3}{c}{LT Measured Parameters}\\
$F_\mathrm{ecl}$ (mmag) & 0.44 $\pm$ 0.21 & B \\
$\sigma_\mathrm{w}$ & (7.7 $\pm$ 3.6) x 10$^{-4}$ & B \\
$\sigma_\mathrm{r}$ & (2.71 $\pm$ 0.65) x 10$^{-4}$ & B \\
$T_\mathrm{B}$ (K) &  $2995^{+200}_{-280}$  & B \\
  \hline
 \end{tabular}
\end{table}

\citet{2012ApJ...760..140C} and \citet{2013MNRAS.428..182B} detected an elongated companion candidate 1$.\!\!^{\prime\prime}$06 from WASP-12, which further observations by \citet{2013MNRAS.436.2956S} and \citet{2014ApJ...788....2B} confirmed to be a close M-type binary that is gravitationally bound to WASP-12. These were unresolved in the PSF of WASP-12 for both of our observations. Prior to fitting for the eclipse depths for each dataset, we corrected the light curves for the dilution caused by the two stellar companions of WASP-12, as well as the two background stars in our target apertures. A detailed description of how this correction was performed is given in Appendix \ref{appendix}. The corrected light curves are shown in the top panels of Figures \ref{INT} and \ref{LT}.

To extract the eclipse depths, we fit each light curve with an eclipse model using a Markov Chain Monte Carlo (MCMC) method with the Metropolis--Hastings algorithm. We used a Mandel and Agol transit model \citep{2002ApJ...580L.171M} to model the eclipse, with both limb darkening coefficients set to zero. Table \ref{tab:parameters} displays the stellar, planetary and orbital parameters of the WASP-12A system from \citet{2009ApJ...693.1920H}\footnote{Whilst updated, more precise values for these parameters are given in \citet{2017AJ....153...78C}, we adopted the values from \citet{2009ApJ...693.1920H} to allow direct comparison to most of the other eclipse measurements for WASP-12b.} that were fixed in the MCMC (including the mid-transit time). To quantify the noise associated with the residual flux, we used a wavelet likelihood function \citep{2009ApJ...704...51C}. This parameterises the noise in the light curve as a sum of white-noise ($\sigma_\mathrm{w}$), and red-noise ($\sigma_\mathrm{r}$) with power spectral density $1/f^\gamma$.

The model used to fit the normalised light curves included components associated with the eclipse depth, a linear function of sky background, $\sigma_\mathrm{w}$ and $\sigma_\mathrm{r}$ (spectral index $\gamma$ was fixed to 1), which were allowed to vary with each step. For the LT model, a linear function of airmass was also included. We ran an MCMC chain of 10$^6$ steps to get the best fit values for each parameter, and used the Gelman--Rubin criterion \citep{1992StaSc...7..457G} to verify convergence. After the first run, all points over 3$\sigma$ away from the best fit model were discarded (these points are shown as grey asterisks in Figures \ref{INT} and \ref{LT}) and another MCMC chain of 10$^6$ steps was run to determine the final best fit values for each parameter.

The best fit models for each dataset, with depths of 0.97 $\pm$ 0.14 mmag for the INT light curve and 0.44 $\pm$ 0.21 mmag for the LT light curve (which disagree by $\sim 3\sigma$), are shown in red in Figures \ref{INT} and \ref{LT} with the fit parameters shown in Table \ref{tab:parameters}.
\section{Discussion}
\label{sec:discussion}

\begin{figure*}
\centering
\includegraphics{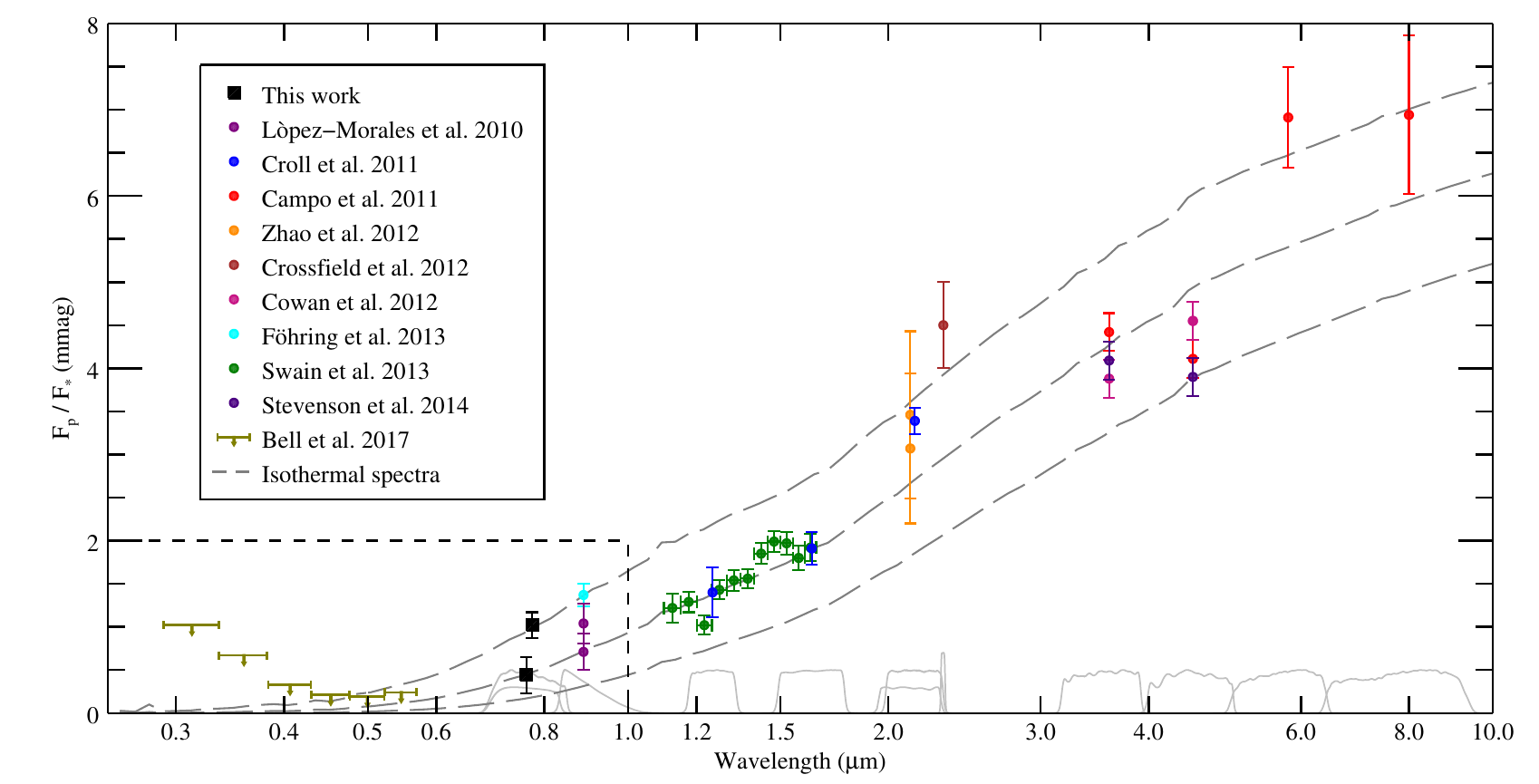}
\caption{Data points show reported eclipse depths of WASP-12b across the NUV, optical and NIR. The depths displayed from \citet{2010ApJ...716L..36L} and \citet{2012ApJ...744..122Z} are the depths associated with each of their individual eclipse observations, not the combined depths that are more commonly used. For the \textit{HST} spectrophotometry from \citet{2017ApJ...847L...2B} and \citet{2013Icar..225..432S}, the horizontal bars denote the width of the bins. The solid grey lines show response functions associated with each of the photometric eclipse measurements. Points marked with a downward arrow denote 2$\sigma$ upper limits on the eclipse depth \citep{2017ApJ...847L...2B}. Grey dashed lines show the eclipse depths associated with WASP-12b dayside emission modelled by blackbody spectra with $T_\mathrm{B}$ = 2600 K, 3000 K and 3400 K. A close-up of optical and NUV wavelengths (shown by the dashed box) in shown in Figure \ref{closeup}.}\label{models}
\end{figure*}

\subsection{Collective analysis of reported WASP-12b eclipse depths}

To interpret our results, we analyse them jointly with all other published secondary eclipse detections of WASP-12b. This is complicated by the fact that WASP-12 (WASP-12A, more strictly) has two M-type companions, and that the previous eclipse measurements were published at different stages in the detection and categorisation of these companions \citep{2012ApJ...760..140C,2013MNRAS.428..182B,2013MNRAS.436.2956S,2014ApJ...788....2B}. Therefore, we perform our own homogeneous correction on all of the uncorrected eclipse depths reported in each paper. This is described in Appendix \ref{appendix}, with the uncorrected and corrected eclipse depths and dilution factors listed in Table \ref{tab:depths}. To keep the methods of reduction and analysis as consistent as possible, we use the depths reported in \citet{2014ApJ...791...36S} for all HST/WFC3 and Spitzer/IRAC data, but differentiate between them in our discussion by referring to the original studies by \citet{2011ApJ...727..125C,2012ApJ...747...82C} and \citet{2013Icar..225..432S}.

\begin{figure}
\includegraphics{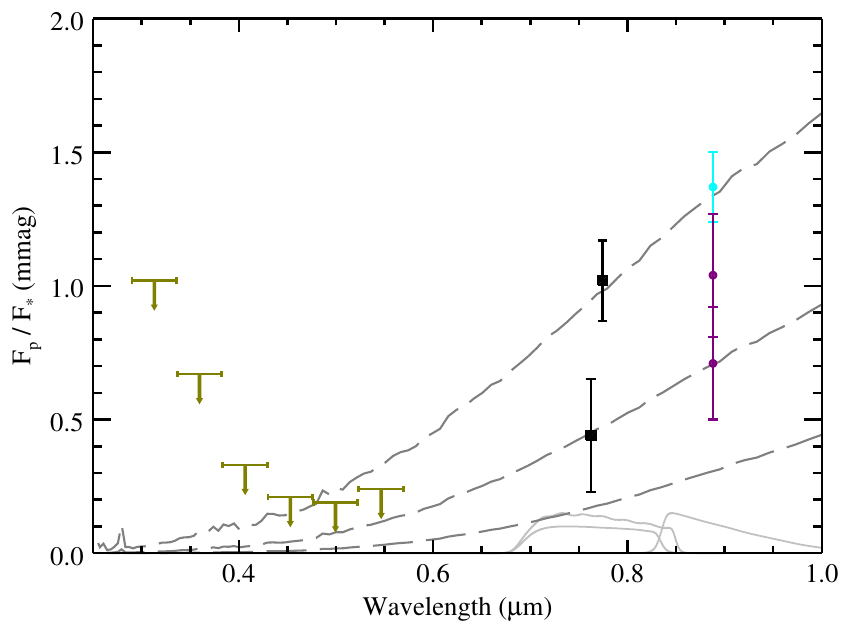}
\caption{A close-up of optical and NUV wavelengths shown in Figure \ref{models}.}\label{closeup}
\end{figure}

To study how well measurements agree in each photometric band, we consider the measured eclipse depths associated with individual observations separately, rather than combining them all into one single depth. These are shown in Figures \ref{models} and \ref{closeup}, along with three blackbody spectra with brightness temperatures ($T_\mathrm{B}$) of 2600 K, 3000 K and 3400 K. We use these three temperatures as benchmarks with which to compare the implied dayside temperatures of each eclipse measurement. For WASP-12, we used a Kurucz stellar model \citep{2004astro.ph..5087C} for a 6250 K star with $\mathrm{log}(g)$ = 4.5 and [Fe/H] = 0.2 \citep[based on values from][shown in Table \ref{tab:parameters}]{2009ApJ...693.1920H}. Whilst the temperature-pressure profile of the WASP-12b dayside is not necessarily isothermal at the altitudes probed by the full range of wavelengths we are considering, it is convenient for us to consider each eclipse in terms of its brightness temperature (which are also listed in Table \ref{tab:depths} for each observation) to compare measurements at different wavelengths. 

The three $z^\prime$-band eclipse measurements exhibit disagreement on a similar scale to our $i^\prime$-band results. The two depths from \citet{2010ApJ...716L..36L} are 1.04 $\pm$ 0.23 mmag from an observation in February 2009 and 0.71 $\pm$ 0.21 mmag from an observation in October 2009, whereas the depth from \citet{2013MNRAS.435.2268F} is 1.37 $\pm$ 0.13 mmag from an observation in January 2010. The shallowest and deepest depths are discrepant by over 3$\sigma$, and were observed just over two months apart. The brightness temperatures of both the shallowest $z^\prime$-band eclipse and our LT eclipse are very close to 3000 K, and the brightness temperatures from \citet{2013MNRAS.435.2268F} and our INT eclipse are both very close to 3400 K.


Whilst the HST/STIS spectrophotometry from \citet{2017ApJ...847L...2B} does not strongly constrain the thermal emission in the three bluest wavelength bins, the three redder bins favour models with lower thermal emission, and rule out $T_\mathrm{B}$ = 3400 K at $>$2$\sigma$ for the 476-523 and 523-570 nm bins and $>$1$\sigma$ for the 430-476 nm bin. The HST/WFC3 spectrophotometric depths from \citet{2013Icar..225..432S} agree well with the overlapping broadband $J$- and $H$-band depths from \citet{2011AJ....141...30C}, and the corresponding brightness temperatures of the majority of the bins are within 1$\sigma$ of 3000 K. We note that the WFC3 data were acquired from the observation of a single secondary eclipse, and the $J$- and $H$-band eclipse observations from \citet{2011AJ....141...30C} occurred on consecutive nights. 

The $K_S$-band observation from \citet{2011AJ....141...30C} results in a brightness temperature of 3294 $\pm$ 62 K. The two $K_S$-band depths reported by \citet{2012ApJ...744..122Z} were both based on observations of partial secondary eclipses on consecutive nights and the large errors associated with each of the depths result in brightness temperatures that are consistent with both 3000 K and 3400 K. The narrowband measurement centred on 2.315 $\mathrm{\mu} m$ by \citet{2012ApJ...760..140C} results in a brightness temperature of 3647$^{+200}_{-210}$ K: over 1$\sigma$ above 3400 K. It should be noted that this depth was extracted from observations with a baseline of a little over 30 minutes. Whilst these results are more in line with hotter dayside temperatures, we also note that $K$- and $K_S$-band observations of other exoplanets have routinely yielded deeper than expected secondary eclipses \citep[e.g.][]{2010MNRAS.404L.114G,2011A&A...528A..49D}.

Although the three secondary eclipses detected using the 3.6 $\mathrm{\mu}$m channel of Spitzer/IRAC all have brightness temperatures within 1.5$\sigma$ of 3000 K, the high precision achieved in each of the measurements leads to a 2.4$\sigma$ disagreement between the depths from the observations by \citet{2011ApJ...727..125C} and \citet{2012ApJ...747...82C}. These observations occurred over the course of 2 years and were separated by at least six months. The four secondary eclipses observed with the 4.5 $\mathrm{\mu}$m IRAC channel tell a similar story, with a 2.9$\sigma$ disagreement between the depths from the observations by \citet{2012ApJ...747...82C} and \citet{2014ApJ...791...36S}. These exhibit some of the lowest brightness temperatures of any band, with the observation by \citet{2014ApJ...791...36S} in agreement with 2600 K and the two eclipse depths from the \citet{2012ApJ...747...82C} phase curve observation within 1.5$\sigma$ of 3000 K. However, various models have used \ce{CO} absorption to explain the shallow depths reported in this channel \citep[e.g.][]{2012ApJ...758...36M,2014ApJ...791...36S}. The depths extracted from the observations in the 5.8 and 8.0 $\mathrm{\mu}$m IRAC channels by \citet{2011ApJ...727..125C} result in some of the highest brightness temperatures for any WASP-12b eclipse observation, and are both consistent with 3400 K. It should be noted that the eclipse depths originally reported in \citet{2011ApJ...727..125C} for the 3.6 and 5.8 $\mathrm{\mu}$m come from one eclipse observation, and the 4.5 and 8.0 $\mathrm{\mu}$m depths come from another eclipse observation. The simultaneous observations in the shorter wavelength channels both yield brightness temperatures that are in disagreement with the longer wavelength observations by $\sim$2$\sigma$.

We can conclude that disagreements in secondary eclipse depths for WASP-12b occur at $>$2$\sigma$ in every photometric band where multiple observations have been conducted (apart from in the $K_S$ band, where two of the three depths had large errors associated with them). However, the disagreements in the 3.6 and 4.5 $\mathrm{\mu}$m channels could be explained by a systematic under-reporting of uncertainties, as the associated depths were reported at high significance and cover a much narrower range of brightness temperatures than results in the $i^\prime$- and $z^\prime$-bands. No published studies to date have been able to formulate models that fit reported depths in all bands across the IR. Along with the disagreements at shorter wavelengths, this presents challenges in interpreting the physical significance of these data. From our analysis, whilst it is certainly likely that sizeable systematic errors bias some of the results discussed above, we conclude that we cannot rule out the possibility that the observed disagreements are caused by the thermal emission varying by an average of a few hundred Kelvin in the dayside of WASP-12b.

\subsection{Potential sources of systematic error}

It is highly plausible that the observed disagreements in measured eclipse depths could arise due to systematic errors being introduced into the time-series at various stages of data acquisition, reduction and analysis. In this section, we will qualitatively discuss some of the most likely sources of these.

In terms of the data acquisition, secondary eclipses of WASP-12b have been observed using a wide selection of instruments on different telescopes, each of which have their own associated characteristic properties. There are variations in sensitivity between pixels on a detector, that flat-fielding does not completely remove. Imperfect telescope guiding leads to the target drifting across the detector, which results in differences in the sensitivities of the pixels that the PSFs are occupying. This could introduce sizeable time-correlated systematics into the resulting light curve. Systematics associated with weather and atmospheric conditions on Earth can be introduced due to secondary eclipse observations occurring at different times and at different locations. This could introduce large systematics into the resulting light curves, especially if the weather is variable on a short timescale within a given observation. 

The $i^\prime$- and $z^\prime$-bands contain numerous telluric emission and absorption lines, both of which can be the source of systematic errors. Telluric emission lines due to \ce{OH-} radicals cause the sizeable fringing patterns on the detector described in section \ref{sec:observations}, which, if not robustly corrected for, can introduce their own systematic errors. This is especially true if a target or comparison star crosses a fringe during time-series observations. There is also significant telluric \ce{H2O} absorption at these wavelengths, although most observations are relatively robust against non-uniform water content in the atmosphere due to the small angular separations between the target and comparison stars. Contamination from satellites, the Moon and cosmic rays can cause spikes in the observed flux for small sections of the time-series if the frames are not corrected or removed from the analysis.

Once the data have been acquired, differences in methods of reduction--including bias subtraction, flat-fielding and other more specialised steps--can introduce systematic errors. There are many methods used to fit eclipse light curves, with the size of the eclipse and associated errors often highly dependent on which baseline model is used. This issue can arise when two or more different baseline models that result in significantly different eclipse depths are judged to model the residual flux equally well. Observations of partial eclipses, or observations where relatively little out-of-eclipse baseline is observed, are particularly susceptible to this problem. Despite the high significance at which many secondary eclipse depths are detected, many of the systematic errors discussed will not manifest themselves in the reported errors. The fact that the biggest disagreement in eclipse depths for WASP-12b occur in the shortest wavelengths could be indicative of the challenges associated with trying to accurately measure the small signal sizes associated with the thermal emission at these wavelengths.


Whilst a detailed discussion of individual disagreements is beyond the scope of this paper, a common source of such disagreements is differences in methods of reduction and analysis, as reanalysis of datasets by different groups often rules out previously stated conclusions \citep[e.g.][]{2011MNRAS.411.2199G,2014MNRAS.444.3632H,2019MNRAS.482.2065E}. It is considerably harder to quantify and correct for any systematics associated with instruments and weather conditions.

\subsection{Variability in dayside thermal emission}

Another potential explanation of the discrepancies in reported secondary eclipse depths is that they arise due to atmospheric variability in the thermal emission properties of the WASP-12b dayside itself. The most compelling observational evidence for variability in the atmosphere of a hot Jupiter comes from HAT-P-7b. \citet{2016NatAs...1E...4A} fit for the planetary brightness amplitude $A_p$, secondary eclipse depth $F_\mathrm{ecl}$ and peak brightness offset $\Theta$ in the phase curves of four years of \textit{Kepler} photometry of HAT-P-7b. Whilst no significant variation was detected for $A_p$ and $F_\mathrm{ecl}$, $\Theta$ varied between -0.086 and 0.143 in phase (occurring both before and after the secondary eclipse) in a highly time-correlated manner, on a timescale of months. This can be interpreted as the brightest spot on the HAT-P-7b dayside shifting about the substellar point. \citet{2016NatAs...1E...4A} put forward the explanation of changing wind speeds, which transport clouds that form on the nightside of HAT-P-7b different distances onto the dayside before they evaporate, causing temporal variability in the thermal and reflective properties of the dayside. No significant variability in $\Theta$ was observed in the \textit{Kepler} light curves of Kepler-13Ab; the only other planet within the \textit{Kepler} field suitable for a similar study to be conducted \citep{2018AJ....155...13H}.

\citet{2013MNRAS.435.2268F} suggested that local variations in surface brightness could explain the differences in reported eclipse depths for WASP-12b. In this case, fluctuations of $\sim$450-750 K relative to a background of 3000 K for 10-20\% of the surface would be sufficient to account for the disagreement. \citet{2014ApJ...791...36S} cast doubt on this idea on the basis of the short ($\sim$ 2 hour) out-of-eclipse baseline from which the deep eclipse reported by \citet{2013MNRAS.435.2268F} was measured. Whilst the small depth extracted from our LT observation was measured from an even shorter out-of-eclipse baseline (1.7 hours), we note that the $T_\mathrm{B}$ associated with our INT observation (which had $\sim$3.9 hours of out-of-eclipse baseline) is in strong agreement with that of \citet{2013MNRAS.435.2268F}. This suggests that these disagreements are not solely due to poorly-constrained baseline models biasing the reported eclipse depths.

WASP-19b is the only other exoplanet for which multiple $z^\prime$-band eclipse depths have been reported. Whilst the depths reported by \citet{2012ApJS..201...36B} and \citet{2013ApJ...774..118Z} are in good agreement, \citet{2013A&A...552A...2L} reported a shallower depth in $\sim$2$\sigma$ disagreement, on the basis of 10 observations over the course of 2 years. The $i^\prime$- and $z^\prime$-bands probe planets bluewards of the peak of their thermal emission, so secondary eclipse measurements in these bands are very sensitive to temperature variations. Further strategic observations of ultra-hot Jupiters are required to discriminate between whether these disagreements in red optical secondary eclipse depths are due to systematics or variations in the dayside thermal emission.
%

\section{Conclusions}
\label{sec:conclusions}

We have presented the reduction and analyses of two $i^\prime$-band secondary eclipse observations of WASP-12b. Much like similar observations at $z^\prime$-band wavelengths, the measured depths disagree by $\sim 3\sigma$. This presents challenges in interpreting the physical cause of these results, with the depths analysed in isolation indicating markedly different things about dayside temperature and presence of TiO. This disagreement could arise either due to systematic errors manifest in the light curves from which these depths were extracted, or genuine variability in the dayside atmosphere of WASP-12b itself. 

Our results demonstrate that 2m-class telescopes are sufficiently large to yield results with the requisite precision to detect significant disagreement between individual eclipse depths at red-optical wavelengths for WASP-12b. Homogeneous reanalysis of published datasets and repeat measurements with different instrumentation \citep[e.g.][]{2014ApJ...791...36S,2017MNRAS.467.4591G,2018A&A...620A.142A} are important methods to test whether such disparities arise due to different methods in reduction and analysis. However, the main stumbling block when trying to discriminate between these scenarios is the low statistics on which any conclusions must be based. Future strategies that involve exploiting the reduced time-pressure on many 2m-class telescopes could allow many secondary eclipses of ultra-short period exoplanets to be observed in quick succession. The ability to observe a sizeable proportion of these observations (or observations of a similar nature) would facilitate the identification of outlying eclipse depths. The strategy of conducting many observations in quick succession would also allow a search for any time-correlation in the eclipse depths to be identified, on a timescale similar to the peak flux offset variability observed by \citet{2016NatAs...1E...4A} for HAT-P-7b. Such a study could be supplemented by observations of the same secondary eclipse by two or more telescopes. This would put to test the very technique of ground-based photometry to observe the secondary eclipses of exoplanets, by investigating whether the same methods of reduction and analysis produce results that agree.

Although there were limited suitable targets within the \textit{Kepler} field to search for variability in transiting exoplanets, the launch of the \textit{Transiting Exoplanet Survey Satellite} \citep[\textit{TESS};][]{2015JATIS...1a4003R} presents the opportunity to continue this search for a wide selection of targets across the night sky. Hot Jupiters within its continuous viewing zones (most notably WASP-100) will be excellent targets to perform similar studies to that of \citet{2016NatAs...1E...4A}. Outside these zones, the 27+ days of \textit{TESS} photometry may be sufficient to identify evidence of atmospheric variability for hot Jupiters with large secondary eclipse depths orbiting bright stars (such as KELT-9b and WASP-33b). Recently, \citet{2018arXiv181106020S} demonstrated the capability of this technique to detect the secondary eclipse and phase modulation of WASP-18b based on 54 days of \textit{TESS} photometry. Finally, the upcoming launch of the \textit{CHaracterising ExOPlanets Satellite} \citep[\textit{CHEOPS};][]{2014SPIE.9143E..2JF} will provide further opportunity to observe targeted full-phase curves of any object in the night sky.

\section*{Acknowledgements}

Special thanks go to Kevin Stevenson for his assistance in the preparation of this manuscript. We thank the anonymous referee for their meticulous feedback, which materially improved our study in many regards. We are also grateful to the staff of the INT and Robert Smith of the LT for their assistance with the described observations. The WFC photometry was obtained as part of programme I/2016B/P6. The INT is operated on the island of La Palma by the Isaac Newton Group of Telescopes in the Spanish Observatorio del Roque de los Muchachos of the Instituto de Astrof{\'i}sica de Canarias. IO:O photometry was obtained as part of programme PL18A05. The Liverpool Telescope is operated on the island of La Palma by Liverpool John Moores University in the Spanish Observatorio del Roque de los Muchachos of the Instituto de Astrof{\'i}sica de Canarias with financial support from the UK Science and Technology Facilities Council. This publication makes use of data products from the Two Micron All Sky Survey, which is a joint project of the University of Massachusetts and the Infrared Processing and Analysis Center/California Institute of Technology, funded by the National Aeronautics and Space Administration and the National Science Foundation.

MJH \& SRM acknowledge funding from the Northern Ireland Department for the Economy. CAW acknowledges support from STFC grant ST/P000312/1. NPG acknowledges support from the Royal Society in the form of a University Research Fellowship.




\bibliographystyle{mnras}
\bibliography{WASP12_i_MNRAS_citations} 




\appendix

\section{Correcting for contaminating stars}
\label{appendix}

Where applicable, we corrected for contaminating flux from the two stellar companions of WASP-12, as well as two nearby background stars. As well as correcting for this contamination for our data (using the values shown in Table \ref{tab:contaminants}), we performed a homogeneous correction for all reported eclipse depths of WASP-12b, which facilitates the direct comparison of different eclipse depths. The dilution factors, uncorrected and corrected eclipse depths and resulting brightness temperatures are shown for each reported eclipse observation in Table \ref{tab:depths}. The calculated dilution factors and corrected eclipse depths are broadly within 1$\sigma$ of the equivalent values from studies that performed similar corrections for WASP-12b eclipse data \citep{2012ApJ...760..140C,2014AJ....147..161S,2014ApJ...791...36S}

For the two stellar companions of WASP-12, we used the spectral classification of M0V from \citet{2012ApJ...760..140C} and \citet{2013MNRAS.436.2956S}. Two other nearby stars (2M063033 and 2M063032) were inside the target apertures for the reduction of two of our datasets, as well as the $z^\prime$-band observation from \citet{2013MNRAS.435.2268F}. We estimated the spectral types of the these two stars using their 2MASS $J$, $H$ and $K_s$ magnitudes from \citet{2006AJ....131.1163S}, as well as the $J-H$ and $H-K_S$ values from \citet{2013ApJS..208....9P}. We modelled the spectra of WASP-12 and each of the contaminating stars using Kurucz stellar models \citep{2004astro.ph..5087C}. We then integrated the spectra over the 2MASS $J$ filter and used the 2MASS $J$ magnitudes for each star to calculate dilution factors for each photometric band and spectral wavelength bin for which eclipse depths have been reported for WASP-12b. When analysing our own data, we removed the flux associated with the contaminating stars for each dataset by normalising the light curves so that the median of the out-of-eclipse baseline was equal to 1, subtracting the dilution factor for the associated photometric band (shown in Table \ref{tab:contaminants}) and repeating the normalisation. For all of the other previous reported eclipse depths, we made the dilution correction using the equation
\begin{equation}
\delta_\mathrm{corr}\left ( \lambda  \right )=[1+g(\beta,\lambda)\alpha_\mathrm{comp}(\lambda)]\delta_\mathrm{meas}(\lambda),
\end{equation}
where $\delta_\mathrm{corr}\left ( \lambda  \right )$ and $\delta_\mathrm{meas}\left ( \lambda  \right )$ respectively are the corrected and uncorrected eclipse depths, $g(\beta,\lambda)$ is the fraction of the flux of the contaminating star inside the target aperture of size $\beta$ and $\alpha_\mathrm{comp}(\lambda)$ is the dilution factor \citep{2014ApJ...791...36S}. 

The two stellar companions are unresolved from WASP-12 for all datasets, except the HST/STIS spectrophotometry from \citet{2017ApJ...847L...2B} which required no correction. We set $g(\beta,\lambda)=1$ for bandpasses and spectral wavelength bins apart from the Spitzer/IRAC data, where we used values from \citet{2014AJ....147..161S} and \citet{2014ApJ...791...36S}.

\begin{table}
 \caption{The flux contribution (expressed as a \% of the flux of WASP-12) from contaminating stars inside the target aperture (see section \ref{sec:analysis} for details). *The spectral determination of WASP-12B and WASP-12C is from \citet{2012ApJ...760..140C} and \citet{2013MNRAS.436.2956S}.}
 \label{tab:contaminants}
 \begin{tabular}{lccc}
  \hline
  Object & Sp. Type & INT flux & LT flux\\
   & & (\%) & (\%)\\
  \hline
  WASP-12B & M0V* & 1.417 $\pm$ 0.093 & 1.398 $\pm$ 0.091 \\
  WASP-12C & M0V* & 1.328 $\pm$ 0.087 & 1.311 $\pm$ 0.086 \\
  2M063033 & $\approx$K2V & 0.848 $\pm$ 0.066 & 0.845 $\pm$ 0.066 \\
  2M063032 & $\approx$K5V & 0.174 $\pm$ 0.033 & 0.173 $\pm$ 0.033\\
  \hline
  & Total & 3.77 $\pm$ 0.28 & 3.73 $\pm$ 0.27 \\
  \hline
 \end{tabular}
\end{table}

\bsp	

\begin{landscape}
\begin{table}
 \caption{The measured depths, along with our calculated dilution factors, corrected depths and brightness temperatures associated with all of the previously reported secondary eclipse observations of WASP-12b.}
 \label{tab:depths}
 \begin{tabular}{cccccccc}
  \hline
Filter/Grism &  Telescope/Instrument & Central Wavelength & Source & Dilution Factor & Uncorrected Depth & Corrected Depth & $T_\mathrm{B}$\\
 & & ($\mathrm{\mu}$m) &  & (\%) & (mmag) & (mmag) & (K) \\
\hline
\multirow{2}{*}{$i^\prime$} & INT/WFC & 0.774&This work  & 3.77 $\pm$ 0.28 & 0.97 $\pm$ 0.14 & N/A & $3412^{+85}_{-93}$ \\ \cline{2-8}
 & LT/IO:O & 0.763 &This work & 3.73 $\pm$ 0.27  & 0.44 $\pm$ 0.21 & N/A & $2995^{+200}_{-280}$\\
  \hline
\multirow{3}{*}{$z'$} &  \multirow{2}{*}{ARC 3.5m/SPICam} & \multirow{2}{*}{0.888} & \multirow{2}{*}{\citet{2010ApJ...716L..36L}}& \multirow{2}{*}{4.10 $\pm$ 0.26} & 0.68 $\pm$ 0.21 & 0.71 $\pm$ 0.21 &  $2986^{+150}_{-180}$\\ 
 & & && & 1.00 $\pm$ 0.23 & 1.04 $\pm$ 0.23 & $3215^{+130}_{-150}$ \\ \cline{2-8}
&   WHT/Ultracam &0.888 & \citet{2013MNRAS.435.2268F} & 5.24 $\pm$ 0.38 & 1.30 $\pm$ 0.13  & 1.37 $\pm$ 0.13 & $3398^{+68}_{-72}$ \\
\hline
$J$ & CFHT/WIRCam & 1.25 & \citet{2011AJ....141...30C} & 6.58 $\pm$ 0.44 & 1.31 $\pm$ 0.28 & 1.40 $\pm$ 0.29 & $3005^{+150}_{-170}$ \\
  \hline
$H$ & CFHT/WIRCam & 1.63 & \citet{2011AJ....141...30C} & 8.74 $\pm$ 0.0.57 & 1.76 $\pm$ 0.18 & 1.91 $\pm$ 0.19 & $3008^{+95}_{-99}$ \\
  \hline
\multirow{3}{*}{$K_S$} & CFHT/WIRCam & 2.15 & \citet{2011AJ....141...30C} &9.70 $\pm$ 0.63 & 3.09 $\pm$ 0.13 & 3.39 $\pm$ 0.15 & 3294 $\pm$ 62 \\ \cline{2-8}
& \multirow{2}{*}{Hiltner/TIFKAM} & \multirow{2}{*}{2.12} & \multirow{2}{*}{\citet{2012ApJ...744..122Z}} & \multirow{2}{*}{9.58 $\pm$ 0.62} & 2.81 $\pm$ 0.85& 3.07 $\pm$ 0.87 & $3176^{+360}_{-390}$ \\
& & & && 3.16 $\pm$ 0.95 & 3.46 $\pm$ 0.97 & $3336^{+390}_{-420}$ \\ \hline
NB2315 & Subaru/MOIRCS & 2.315 & \citet{2012ApJ...760..140C} & 10.50 $\pm$ 0.69 & 4.1 $\pm$ 0.5 & 4.5 $\pm$ 0.5 & $3647^{+200}_{-210}$ \\ \hline
\multirow{3}{*}{IRAC CH1} & \multirow{3}{*}{Spitzer/IRAC} & \multirow{3}{*}{3.6} & \citet{2011ApJ...727..125C} & 10.85 $\pm$ 0.73 & 4.1 $\pm$ 0.2 & 4.42 $\pm$ 0.22 & $3081^{+85}_{-86}$ \\ \cline{4-8}
&  & & \citet{2012ApJ...747...82C} & 10.85 $\pm$ 0.73 &3.6 $\pm$ 0.2 & 3.88 $\pm$ 0.22 & $2871^{+86}_{-87}$ \\ \cline{4-8}
&& & \citet{2014ApJ...791...36S} & 10.85 $\pm$ 0.73 &3.8 $\pm$ 0.2 & 4.09 $\pm$ 0.22 & $2955^{+85}_{-86}$ \\ \hline
\multirow{4}{*}{IRAC CH2} & \multirow{4}{*}{Spitzer/IRAC} & \multirow{4}{*}{4.5} & \citet{2011ApJ...727..125C} & 11.87 $\pm$ 0.78 & 3.8 $\pm$ 0.2 & 4.11 $\pm$ 0.22 & 2711 $\pm$ 85 \\ \cline{4-8}
&  & & \multirow{2}{*}{\citet{2012ApJ...747...82C}} & \multirow{2}{*}{11.87 $\pm$ 0.78} & 4.2 $\pm$ 0.2 & 4.55 $\pm$ 0.22 & $2877^{+84}_{-85}$  \\ 
&& & & & 4.2 $\pm$ 0.2 & 4.55 $\pm$ 0.22 & $2877^{+84}_{-85}$  \\  \cline{4-8}
&& & \citet{2014ApJ...791...36S} & 11.87 $\pm$ 0.78 & 3.6 $\pm$ 0.2 & 3.90 $\pm$ 0.22 & 2627 $\pm$ 85\\ \hline
IRAC CH3 & Spitzer/IRAC & 5.8 &  \citet{2011ApJ...727..125C} & 11.65 $\pm$ 0.76 & 6.29 $\pm$ 0.54 & 6.91 $\pm$ 0.58 & 3577 $\pm$ 220 \\ \hline
IRAC CH4 & Spitzer/IRAC & 8.0 & \citet{2011ApJ...727..125C} & 11.55 $\pm$ 0.76 & 6.36 $\pm$ 0.88 & 6.94 $\pm$ 0.92 & 3399 $\pm$ 340 \\ \hline
\multirow{6}{*}{G430L} &  \multirow{6}{*}{HST/STIS} & 0.313 & \multirow{6}{*}{\citet{2017ApJ...847L...2B}} & 0 & $<$ 1.02 (2$\sigma$) & N/A &$<$ 4570 (2$\sigma$) \\
& & 0.360 & & 0 & $<$ 0.67 (2$\sigma$) & N/A & $<$ 4190 (2$\sigma$) \\
& & 0.407 & & 0 & $<$ 0.33 (2$\sigma$) & N/A & $<$ 3810 (2$\sigma$) \\
& & 0.453 & & 0 & $<$ 0.21 (2$\sigma$) & N/A & $<$ 3520 (2$\sigma$) \\
& & 0.500 & & 0 & $<$ 0.19 (2$\sigma$) & N/A & $<$ 3330 (2$\sigma$) \\
& & 0.547 & & 0 & $<$ 0.24 (2$\sigma$) & N/A & $<$ 3270 (2$\sigma$) \\ \hline
\multirow{11}{*}{G141} &  \multirow{11}{*}{HST/WFC3} & 1.125 & \multirow{11}{*}{\citet{2013Icar..225..432S}} & 5.79 $\pm$ 0.38 & 1.16 $\pm$ 0.17 &1.22 $\pm$ 0.17 &$3030^{+95}_{-103}$  \\
& & 1.175 & & 6.10 $\pm$ 0.41 & 1.22 $\pm$ 0.12 & 1.29 $\pm$ 0.12 & $3019^{+66}_{-70}$ \\
& & 1.225 & & 6.43 $\pm$ 0.42 & 0.96 $\pm$ 0.11 & 1.02 $\pm$ 0.11 & $2811^{+72}_{-76}$ \\
& & 1.275 & & 6.78 $\pm$ 0.44 & 1.34 $\pm$ 0.11 & 1.43 $\pm$ 0.11 & $3005^{+59}_{-62}$ \\
& & 1.325 & & 6.83 $\pm$ 0.45 & 1.44 $\pm$ 0.11 & 1.54 $\pm$ 0.12 & $3035^{+61}_{-63}$\\
& & 1.375 & & 6.81 $\pm$ 0.45 & 1.46 $\pm$ 0.11 & 1.56 $\pm$ 0.11 & $3010^{+60}_{-62}$\\
& & 1.425 & & 6.91 $\pm$ 0.45 & 1.73 $\pm$ 0.11 & 1.85 $\pm$ 0.12 & $3120^{+58}_{-59}$\\
& & 1.475 & & 7.19 $\pm$ 0.47 & 1.85 $\pm$ 0.12 & 1.99 $\pm$ 0.12 & $3156^{+58}_{-60}$ \\
& & 1.525 & & 7.69 $\pm$ 0.50 & 1.83 $\pm$ 0.12 & 1.97 $\pm$ 0.13 & $3121^{+63}_{-65}$ \\
& & 1.575 & & 8.40 $\pm$ 0.55 & 1.66 $\pm$ 0.13 & 1.80 $\pm$ 0.14 & $2996^{+71}_{-73}$\\
& & 1.625 & & 9.12 $\pm$ 0.60 & 1.76 $\pm$ 0.15 & 1.92 $\pm$ 0.16 & $3026^{+82}_{-85}$\\ \hline

 \end{tabular}
\end{table}
\end{landscape}



\label{lastpage}
\end{document}